\documentstyle[preprint,aps,prb]{revtex}
\begin{document}

\draft
\title{Progressive Evolution of Tunneling Characteristics of
In-Situ Fabricated Intrinsic Josephson Junctions in
Bi$_2$Sr$_2$CaCu$_2$O$_{8+\delta}$ Single Crystals}

\author{Yong-Joo Doh, Hu-Jong Lee, and Hyun-Sik Chang}

\address{Department of Physics,
Pohang University of Science and Technology \\
Pohang 790-784, Republic of Korea}

\maketitle

\begin{abstract}
Stacks of a few intrinsic tunnel junctions were
micro-fabricated on the surface of
Bi$_2$Sr$_2$CaCu$_2$O$_{8+\delta}$ single crystals. The number of
junctions in a stack was tailored by progressively increasing the
height of the stack by ion-beam etching, while its tunneling
characteristics were measured in-situ in a vacuum chamber for
temperatures down to $\sim$13 K. Using this in-situ
etching/measurements technique in a single piece of crystal,
we systematically excluded any spurious effects arising from
variations in the junction parameters and made
clear analysis on the following properties of the
surface and inner conducting planes.
First, the tunneling resistance
and the current-voltage curves are scaled by
the surface junction resistance. Second, we confirm that
the reduction in both the gap and the superconducting transition
temperature of the surface conducting plane in contact with a
normal metal is not caused by the variation in the doping level,
but is caused by the proximity contact.
Finally, the main feature of a
junction is not affected by the presence of other junctions in a
stack in a low bias region. \\
\end{abstract}

\newpage
\narrowtext

\section{INTRODUCTION}
Recently both superconducting and normal-state tunneling
characteristics of junctions intrinsically formed in crystals of
extremely anisotropic high-$T_c$ materials such as
Bi$_2$Sr$_2$CaCu$_2$O$_{8+\delta}$ (Bi-2212) and
Tl$_2$Ba$_2$Ca$_2$Cu$_3$O$_{10+\delta}$ (Tl-2223) have attracted
much research attention.
\cite{1,2,3,4,5,6,7,8,9,10,11,12,13,14,15,16,17,18,19,20,21,22,22a,23,24,25,26,37,43}
Previous experimental investigations by many groups have revealed
that the superconducting order parameter in the crystals is
periodically modulated along the $c$-axis and the interlayer
coupling at low-enough temperatures has a Josephson nature.
Existence of such naturally formed intrinsic Josephson junctions
has been directly confirmed by $c$-axis tunneling measurements,
using small-sized stacks or mesa structures on the surface of
single crystals, including current-voltage ($IV$)
characteristics,\cite{2,3,4,5,6,7,8,9,10,11,12} temperature or
magnetic-field dependence of the critical
current,\cite{13,14,15,16,17,18} dynamics of Josephson
vortices,\cite{19,20,21,22,22a} microwave responses,\cite{23,24,25}
and the Josephson microwave emission.\cite{26}

Electron tunneling spectroscopy has been conceived as one of the
most powerful means to determine superconducting gap and its
anisotropy.\cite{27,29,28,30,31,33,32} Compared to other
spectroscopic techniques,\cite{34,35,36} electron tunneling
spectroscopy has a very high energy resolution of the electronic
structure near the Fermi level $E_F$. Tunneling spectroscopy and its
interpretation on high-$T_c$ materials using conventional
techniques such as scanning tunneling spectroscopy, however, have
often been controversial, due to problems of poor surface
characterizations mainly arising from an extremely short coherence
length and uncertainty in the transverse momentum direction of the
particles upon tunneling via a small tip.\cite{28,30} These
problems have been conveniently circumvented by a tunneling study
using intrinsic junctions (IJs) in high-$T_c$ materials.
Especially, tunneling measurements on a small-sized stack,
including only a few intrinsic junctions, provide valuable
information on the nature of the inherent superconducting gap and
the pairing mechanism of unusual symmetry in the materials.

It is unsettled yet, however, whether the individual junctions in
a stack behave independently or are correlated with each other.
Junctions in a stack could be inductively coupled by supercurrents
flowing on thin conducting planes\cite{19,40} or coupled by charging of
thin conducting planes by tenneling electrons.\cite{41} Thus
the plasma oscillations excited in a stack of intrinsic junctions,
for example, show collective modes.\cite{2,24} However, some previous studies
indicate the opposite may be true. For dc bias current applied
along the $c$-axis, coupling between adjacent junctions is small
enough so that each junction can be assumed to be
independent.\cite{5,9} Correlated junction behavior would affect
the usefulness of a stack for device applications. For instance,
correlated junction behavior is useful for high-frequency
oscillator applications but detrimental to voltage standards
applications.\cite{24} Thus, the issue of correlation of the tunneling
characteristics of intrinsic junctions still remains of prime
concern from both academic and device-application points of view.

Tunneling spectroscopy on the surface of high-$T_c$ materials is
widely used to probe the nature of order-parameter symmetry,
mostly using hybrid junctions consisting of conventional and
high-$T_c$ superconducting electrodes.\cite{43,42} In this case,
it is utterly important to make sure that the superconducting
properties of the surface layer, such as the transition
temperature, the gap value, and its order-parameter symmetry, do
not deviate from those of inherent bulk properties.\cite{44}
Recently, surface-sensitive spectroscopy such as anlge-resolved
photoemission spectroscopy,\cite{35} scanning tunneling
spectroscopy,\cite{31,33} break-junction tunneling,\cite{32} and
intrinsic-junction tunneling,\cite{37,38,39} is increasingly used
to investigate the in-plane normal-state conducting state of
high-$T_c$ materials {\it i.e.}, the existence and nature of the
pseudogap, especially as varying the doping level. Since the size
of the pseudogap is known to be sensitive to the doping level in
the underdoped regime,\cite{31,33,32} any possible significant
oxygen deficiency in the surface layer would lead to a very
serious misinterpretation of the intrinsic bulk properties of the
conducting planes. Additionally, our previous study\cite{11}
indicates that depositing noble-metal film such as Au on the
surface of high-$T_c$ materials, often used for surface protection
for electron conduction measurements, causes a suppression of the
surface layer's superconductivity by its proximity contact to
normal metal. In this respect, it is of prime importance to probe
any variation in the doping level of the surface layer from that
of the inner layers with intrinsic bulk properties.

Most studies on the intrinsic tunneling effect in high-$T_c$
superconductors, both in superconducting and normal states, have
been done to date by fabricating a small elevated stack structure
(with the cross-sectional area usually of a few hundreds
$\mu$m$^2$) on the surface of single crystals with a few IJs in
it. The number of the junctions in a stack is varied by changing
the height of the stack during the fabrication process using
ion-beam etching or chemical wet etching. In this way, however,
accurate tailoring of the number of junctions is a difficult task.
To get deeper understanding of the interlayer coupling
characteristics of an individual intrinsic junction as well as the
conducting properties in a plane, both in superconducting and
normal states, careful in-situ control of the number of IJs in a
stack is required while monitoring their tunneling
characteristics.\cite{8}

In this report, we first examine the possibility of
any significant variation of the
doping level in the surface layer of high-$T_c$ single crystals in
contact with normal metals, along with a change in the
superconducting properties. We then discuss the physical
implication of the scaling of tunneling resistance and the
$IV$ curves with respect to surface junction
resistance. We also examine whether the main feature of a junction
is influenced by the presence of other junctions in a stack.
To that end, we focus on the evolution of the
tunneling characteristics of IJs while progressively increasing
the number of junctions in a stack fabricated on the surface of
Bi-2212 single crystals. Ion-beam etching was used in combination
with in-situ cryogenic measurements, similar to the technique
reported in Ref. \onlinecite{8}. The number of IJs in a stack was
increased in sequence by carefully controlling the low-energy
ion-beam etching time with preformed metallic electrodes on the
surface of the stack. Between each etching stage the $c$-axis
tunneling $IV$ characteristics and the temperature dependence of
the $c$-axis tunneling resistance $R_c(T)$ were taken while
varying the substrate temperature from room temperature down to
$\sim 13$ K.

This study directly confirms our previous finding that a weak
junction (WJ) forms at the surface of a Bi-2212 stack\cite{11}
which is in contact with a normal-metal electrode. The in-situ
etching/measurements technique confirms that the suppression of
the gap and the transition temperature in the surface layer of a
stack does not result from the change in the doping level or from
the proximity-induced weak superconductivity in the Bi-O
layer\cite{5,6} which may be surfaced. Rather, the suppression
results presumably from the proximity contact of the surface layer
with $d_{x^2 -y^2}$ symmetry to the Au normal metal. This study
also convincingly demonstrates that the individual junction is
little influenced by the presence of other junctions in a stack in
a low bias region. As to the cause of development of negative
dynamic resistance in the high-bias region of the $IV$ curves with
increasing number of junctions in a stack, we find that the Joule
heating effect, not the nonequilibrium, can be a good candidate
for that cause.

\section{Experiments}
Bi-2212 crystals were grown from a melt using alumina crucibles,
details of which are described elsewhere.\cite{45} Pieces of
single-crystal platelets with typical size $\sim 0.8 \times 0.4
\times 0.03$ mm$^3$ were selected from a mass of the cooled melt.
A platelet was glued onto a sapphire substrate by negative
photoresist, then hard-baked. An optically smooth surface was
prepared by cleaving a platelet of single crystal using Scotch
tape. Right after cleaving, 1000-\AA-thick Au film was deposited
on the surface of the Bi-2212 crystals to protect the surface from
any contamination during stack-fabrication processes.

We then patterned a large `base mesa' of size $\sim 450 \times 15
\times 1 ~\mu$m$^3$ on the crystal surface, using standard
photolithography and Ar-ion-beam etching, with beam voltage and
current density $V_{beam}$ = 300 V and $I_{beam}$ = 0.8 mA/cm$^2$,
respectively. Any residue of burned-out photoresist on the crystal
surface was stripped off by oxygen-plasma etching. Then a
patterned layer of photoresist was placed around the base mesa and
hard-baked to insulate the region, excluding the top surface of
the base mesa. A thick Au film ($\sim$ 8000 \AA) was
deposited,\cite{46} and connection pads from the base mesa were
patterned and ion-beam etched. Thereafter, the electric pads of Au
(totally $\sim$ 9000 \AA) crossing the base mesa acted as masks
for the small stacks in the further in-situ etching process. In
this stage, a thin (less than 300 \AA$~$ in nominal thickness) Au
film was intentionally left on the top surface of the base mesa
between the pads to prevent the inadvertent premature formation of
small stack structures. The remaining Au film would be removed
during the next in-situ measurement process with controlled
ion-beam etching. Figure 1 illustrates the optical photograph of a
specimen after completing the whole processes prior to the in-situ
etching.

For in-situ monitoring of the tunneling characteristics of a small
stack, the specimen was loaded into an ultra-high-vacuum chamber
equipped with a liquid-helium cooling pot as well as an ion-beam
etching system. Ion-beam etching and cryogenic measurements could
be repeated without breaking the vacuum. The remnant 300-\AA-thick
Au film and Bi-2212 crystal underneath were further etched, while
forming four small stacks on the base mesa. The junction size was
defined by the width ($\sim 15$ $\mu$m) of the base mesa and
($\sim 12$ $\mu$m) of the electrical pads overlaid perpendicularly
to the length of the base mesa. In this etching stage we used
sufficiently low ion-beam voltage ($V_{beam}$ = 100 V) and current
density ($I_{beam}$ = 0.16 mA/cm$^2$) to enhance the
controllability of the etching rate. The height of a small stack
was controlled by the total etching time $t_e$. In every 1-2 min
etching, the chamber was pumped out and a stack was characterized
in a three-probe-measurement configuration [see the inset in Fig.
2(b)]. The $R_c (T)$ was measured using a dc method with a
constant bias current of 10 $\mu$A. To reduce the effect of
external noise, low-pass filters were connected to the voltage and
the current leads. A recent report\cite{43} indicates that, using
the similar cleaving technique on Bi-2212 crystals, step
structures with a height of one half unit cell usually forms in
the samples in an area of 30 $\times$ 30 $\mu$m$^2$. We did not
observe any evidence for the existence of a step in our tunneling
results. In this study, we took measurements for three different
stacks, all showing similar features. We present a typical set of
data obtained from one of the specimens.

\section{Results and Discussion}
\subsection{$R_c(T)$ Curves}
Progressive evolution of $R_c (T)$ by increasing the in-situ
etching time $t_e$ up to 30 min is presented in Fig. 2(a).
Increment of the number of junctions in the stack with increasing
$t_e$ was monitored by counting the number of quasiparticle
branches in the corresponding $IV$ curves. Fig. 2(a) reveals that
when we lower the temperature from room temperature down to the
bulk superconducting transition temperature, the $R_c(T)$ curves
show metallic behavior at $t_e \leq 5$ min but gains a slightly
semiconducting behavior at $t_e \geq 6.5$ min where the value of
$R_c$ reaches a maximum at $T_{c,onset}$ ($\simeq$ 92 K). The
value of $R_c$ then starts dropping abruptly and reaches a minimum
at $T_c$ ($\simeq$ 86.5 K). Interestingly, for all the values of
$t_e$, $R_c$ remains finite below $T_c$ and gradually increases
again up to the secondary resistance maximum at
$T_{c,onset}^\prime$. In the range of $T_{c,onset}^\prime < T <
T_c$ the temperature dependence of $R_c$ shows a more pronounced
semiconducting behavior. Below $T_{c,onset}^\prime$ the value of
$R_c$ drops sharply, especially for high values of $t_e$; and
vanishes to 0.1 $\Omega$ at $T_c^\prime$ ($\simeq$ 15 K) and
below.

As $t_e$ increases, $T_{c,onset}^{\prime}$ decreases from $\sim$
36 K to $\sim$ 17 K, as denoted by the dotted guideline in Fig.
2(a). All other characteristic temperatures, $T_{c,onset}$, $T_c$,
and $T_c^{\prime}$, however, remain almost insensitive to the
etching time. The transition width $\Delta T_c^{\prime}$ ($\simeq$
0.3$\sim$5.0 K) near $T_c^{\prime}$ gets sharper with increasing
$t_e$, while $\Delta T_c$ ($\simeq$ 2.3 K) near $T_c$ remains
unchanged. The behavior of $R_c(T)$ near $T_{c,onset}^{\prime}$ is
not only sensitive to $t_e$, but is also highly affected by the
ambient condition. For instance, the peak value of
$R_c(T_{c,onset}^{\prime})$ increases rapidly and $T_c^{\prime}$
is suppressed significantly for RF irradiation of power level as
low as $-30$ dBm (not shown). $R_c(T)$ near $T_{c,onset}^{\prime}$
is so sensitive to the RF noise that the data taken without
filters are greatly affected.

For the sake of clarity we divide the temperature range into three
regions: Region I ($T<T_{c,onset}^{\prime}$), II
($T_{c,onset}^{\prime}<T<T_c$), and III ($T>T_c$), as shown in
Fig. 2(a). The total number of quasiparticle branches obtained
from the $IV$ curves in Region I for the corresponding etching
times are $n=4$ ($t_e$=6.5 min), 4 (8 min), 5 (10 min), 5 (12
min), 6 (15 min), 8 (19 min), 9 (22 min), 10 (26 min), and 12 (30
min).

The finite value of $R_c$ found in the stack in Region II is known
to be always present in a three-terminal
measurement.\cite{3,5,11,17,25} Since the IJs in the stack below
$T_c$ should be in zero-resistance state for a low bias current,
the finite $R_c$ at temperatures in Region II in this
three-terminal configuration is attributed to the $surface$ weak
junction (WJ) consisting of the topmost Cu-O bilayer in the normal
state, and the adjacent inner Cu-O bilayer in the superconducting
state.\cite{11} In other words, the surface Cu-O bilayer
underneath the normal-metal electrode has a suppressed transition
temperature, which is essentially $T_{c,onset}^{\prime}$. Below
$T_c^{\prime}$, the surface junction also becomes Josephson
coupled and the junction resistance subtracts from the value of
$R_c$ for a low current bias.\cite{47} In this region, $R_c$ is
then essentially the contact resistance between the surface Cu-O
bilayer and the Au electrode and its value reduces to as small as
0.1 $\Omega$ (not shown in detail in the figure), which is at
least two orders of magnitude smaller than $R_c$ in other
temperature ranges. We can thus consider $R_c(T_c)$ as the
junction resistance of the surface junction.

We can summarize the above discussion in terms of the tunneling
configuration of the stack and the corresponding $c$-axis
resistance in the following way. In Region I, the stack has
Au/D$^\prime$IDIDID..... configuration, where D (D$^\prime$)
denotes the (suppressed) superconducting layer with $d_{x^2 -
y^2}$ symmetry. $R_c$ in this case is essentially the contact
resistance $R_{CT}$ between Au electrode and D$^\prime$ layer;
$R_c(T)=R_{CT}\lesssim 0.1 ~\Omega$. In Region II, the configuration
becomes Au/N$^\prime$IDIDID....., where N$^\prime$ denotes the
surface layer in its normal state. The corresponding resistance
is\cite{48}
\begin{equation}\label{eq1}
  R_c(T) = R_{CT} + R_{N^{\prime}ID}(T) \simeq R_{N^{\prime}ID}(T)
\end{equation}
\begin{eqnarray}\label{eq2}
  R_{N^{\prime}ID}(T) = \left[ \lim_{V\rightarrow 0} \frac{dI}{dV}
  \right]^{-1}
  = \left\{ \lim_{V\rightarrow 0} \frac{d}{dV}
  \frac{1}{e R_n^{\prime}}
  \int_{-\infty}^{\infty} N(E,\Delta)[f(E)-f(E+eV)] dE \right\}^{-1}
\end{eqnarray}
where $R_{N^{\prime}ID}(T)$ is the quasiparticle tunneling
resistance of the N$^\prime$ID surface junction or WJ,
$R_n^{\prime}$ its junction resistance, and $f(E)$ the Fermi
distribution function. The normalized density of states
$N(E,\Delta)$ of the superconducting Cu-O bilayer with $d_{x^2 -
y^2}$ symmetry can be obtained by averaging over the in-plane
angle in $k$ space as $N(E,\Delta)$ = Re$[(1/2\pi)\int_{0}^{2\pi}
\{ E/ \sqrt{E^2 - [\Delta \cos(2\phi)]^2} \}d\phi$]. Here,
$\Delta$ is the temperature-dependent energy gap of the inner
layers. In Region III, the configuration is
Au/N$^\prime$INININ....., with the tunneling resistance
\begin{eqnarray}\label{eq3}
  R_c(T)&=&R_{CT}+R_{N^{\prime}IN}(T)+m\times R_{NIN}(T)
  \nonumber \\
  &\simeq& R_{N^{\prime}IN}(T)+
  m\times R_{NIN}(T)
\end{eqnarray}
where N denotes the inner Cu-O bilayer in its normal state, $m$ the total
number of IJs (note that $m=n-1$), and $R_{N^{\prime}IN}$ ($R_{NIN}$) stands for
the quasiparticle tunneling resistance of each N$^\prime$IN (NIN) junction.

From Eqs. (\ref{eq1}) and (\ref{eq2}) we note that $R_c(T_c
)\simeq R_{N^{\prime}ID}(T_c)=R_n^{\prime}$ and the value
increases gradually with increasing $t_e$. The values of
$R_c(T_c)$ in Fig. 2(a) are 1.1, 1.2, 1.8, 3.2, 5.2, 6.0, 7.0,
8.3, 9.0, 9.3, and 9.5 $\Omega$ for $t_e$ from 0 up to 30 min [see
Fig. 7(a)]. We presume the variation of $R_c(T_c)$ is due to
shrinking of the junction area as the etching gets longer, the
reason for which will be discussed below.

To study the intrinsic properties of the interlayer coupling or
the conduction in the Cu-O planes, one needs to eliminate the
effect of the variation in the junction area by normalizing
$R_c(T)$ with respect to the junction resistance. Rescaled curve
$R_c(T)/R_c(T_c)$ for each set of curves in Fig. 2(a)
corresponding to $t_e$ equal to 10 min or longer tends to merge
into a single curve at temperatures in Region II [see Fig. 2(b)].
The temperature dependence of the curve is well in accordance with
$R_{N^{\prime}ID}(T)$ expressed in Eq. (\ref{eq2}). Note that in
Eq. (\ref{eq2}) only $R_n^{\prime}$ has $t_e$ dependence, thus
$R_{N^{\prime}ID}/R_n^{\prime}$ or equivalently $R_c(T)/R_c(T_c)$
as plotted in Fig. 2(b) should be independent of $t_e$. The best
fit denoted as the dotted curve is obtained for $\Delta_0 \equiv
\Delta(0)=32.6$ meV with the assumption of the BCS-type
temperature dependence of the gap,\cite{49} $\Delta(T) = \Delta_0
\tanh(\alpha \sqrt{T_c/T - 1})$ with $\alpha=1.45$. Although
$\alpha =1.74$ is valid for the true BCS-type behavior, for this
$d_{x^2 -y^2}$ symmetry $\alpha =1.45$ gives a better fit and is
consistent with the theoretical prediction.\cite{48} The fit turns
out very satisfactory in almost all temperatures in Region II. The
value of $\Delta_0$ is in good agreement with the results of other
studies, obtained from $IV$ curves of the inner IJs with
DID-junction configurations near the liquid-helium
temperature,\cite{3,9,10} the scanning tunneling
spectroscopy,\cite{28,30,31,33} or photo-emission
spectroscopy,\cite{36} using optimally or slightly overdoped
samples.

For comparison, we also draw the best fit curve with an isotropic
$s$-wave gap\cite{50} with $\Delta_0 = 17$ meV (the dashed curve)
which gives, however, only a marginal fit at best near $T_c$. Any
larger $s$-wave gap $\Delta_0$ would result in even a poorer fit
over the range of Region II. Insensitivity of the rescaled
$R_c(T)$ to the number of IJs in a stack in Region II indicates
that the variation of $R_c(T)$, at least for $t_e \geq 10$ min,
was caused by the variation in the junction area with etching.
Note that this insensitivity also implies the number of the
N$^{\prime}$ID junctions does not change with progressive etching;
only one N$^{\prime}$ID junction exists presumably on the surface
of the stack. The deviation of $R_c(T)$ from the merging curve for
$t_e < 10$ min is due to incomplete formation of the stack and
will be discussed below. One notices in Fig. 2(b) that the curves
$R_c(T)/R_c(T_c)$ do not show any merging behavior in Region III,
even for long etching times $t_e > 10$ min. This is consistent
with  the fact  that the $R_c(T)$ in  Region III  include the
tunneling resistances of both number of inner IJs and the surface
WJ, which are scaled by $R_n$ and $R_n^{\prime}$ [$= R_c(T_c)$],
respectively. Thus, $R_c(T)$ in Region III is not scaled by
$R_c(T_c)$ alone.

The inset of Fig. 3 shows more details of $R_c(T)$ curves in
Region III for $t_e = 15$, 19, 22, 26, and 30 min corresponding to
$n=6$, 8, 9, 10, and 12, respectively. Each curve shows   a
pronounced metallic   behavior in the high-temperature region,
which  may imply that  the conducting planes  are in a highly
$overdoped$ regime,\cite{17,38,39} gradually turning into  a
semiconducting behavior  around $T = 170\sim 180$ K  with lowering
temperatures. Increasing $t_e$ tends to  increase $R_c(T)$ in
proportion  as more IJs are  included in the stack. In  Fig. 3  we
plot the single junction  contribution  to the  tunneling
resistivity $\rho_{c,single}$ as converted from the
relation\cite{8} $R _{c,single}(T) = [R_c(T,i+j)- R_c(T,i)]/j$
using the geometric parameters, the junction area $S = 12 \times
15 ~\mu$m$^2$ and its thickness $d = 12$ \AA, for the number of
junctions $i+j=8$, 9, 12, 12 and $i=6$, 6, 8, 6, respectively.
Thus, the curves correspond to  averaging $c$-axis resistivity
over  2, 3, 4, and 6 junctions, respectively. Since $R_c(T_c)$ as
seen in Fig. 2(a) is stabilized for $t_e \gtrsim 15$ min, not much
relative  error is assumed to be involved in this conversion. Note
that  the surface WJ contribution is subtracted from
$\rho_{c,single}$ and it contains only the inner IJ contribution.
One noticeable feature of $\rho_{c,single}$ is that the linear-$T$
behavior is significantly diminished  in the temperature range  of
$T>170$ K, which implies that the linear-$T$ behavior in the inset
of Fig. 3 arises mainly from the  surface WJ. The temperature
dependence of $\rho_{c,single}$ for different etching stages merge
well into a single curve. The dotted curve shows the least $\chi
^2$ fit of the data to the empirical relation
$\rho_c(T)=(a/T)\exp(\Delta^{\star}/T)+bT+c$ as adopted in  Ref.
\onlinecite{39} anticipating  the occurrence of  a pseudogap above
$T_c$,  with the best  fit parameters of $a=146 \pm 8 ~\Omega$cmK,
$\Delta^{\star} = 240 \pm 10$ K, $b=0.0250 \pm 0.008 ~\Omega$cm/K,
and $c = 17.3 \pm 1.0 ~\Omega$cm. The parameter values, especially
the value of $b$, in comparison with Fig. 4 of Ref.
\onlinecite{39} indicate that, although the linear-$T$ dependence
is much reduced, the inner stacks are still in a overdoped regime.
The surface layer, at least with Au protection on it, {\it cannot
be less doped} than the inner layers. Thus, the suppressed
superconductivity in the surface layer  should be attributed to
other causes rather than  its oxygen loss. The  plausible cause is
the  effect of  the proximity  contact of  the surface  layer with
the normal-metal (Au) electrode, as proposed previously.\cite{11}

\subsection{Tunneling $IV$ Curves}
Figure 4 shows the evolution of the quasiparticle branches in the
tunneling $IV$ curves below $T_c^{\prime}$ as the stack  forms
with increasing etching  time $t_e$. For $t_e$  shorter than 5
min, the $IV$ curves show anomalous splitting of quasiparticle
branches. In the  case of $t_e=5$ min as in the inset  of Fig.
4(a), for example, increasing the bias current causes a sudden
drop of voltage to the lower-voltage branch near $V_{drop}\simeq
18$ mV, instead of  a jump to the  next higher-voltage branch as
usually observed in a high-$T_c$
stack.\cite{1,2,3,4,5,6,7,8,9,10,11,12,13,14,15,16,17,18,19,20,21,22}
When the  direction of  the bias  change is  reversed, voltage
jumps around $V_{jump} \simeq 7$ mV to a  neighboring
higher-voltage branch. This anomalous branch  splitting may have
been caused by any remnant Au deposit which was not completely
removed from the surface of the base mesa during the initial stage
of the stack formation.  The remnant deposit appeared to persist
for etching up to  5 min and  partially short  the surface Cu-O
bilayer. We  believe that the  positive temperature coefficient in
the $R_c(T)$ curves above $T_c$ as well as  the deviation of the
curves from the single merging curve in Region II for $t_e \leq 5$
min as seen in Fig. 2(b) was caused by this incomplete formation
of a stack.

In Fig.  4(a), for  $t_e=6.5$  min, one  sees four  branches
develop. As the bias   current is increased voltage jumps to an
adjacent higher-voltage branch at the critical current of  each
branch. Development of this  more normal  behavior of a  stack
indicates  that the regular stack structure started forming at
$t_e=6.5$ min. Even in this case, the critical current $I_c$ (or
the return current $I_r$) for each branch is  much different from
each other.  The $IV$  curve for 8-min  etching shows  similar
behavior, while the critical current in each branch reduced
rapidly (not shown here). The significant discrepancy in $I_c$ (or
$I_r$) most likely resulted from the differences in the junction
area for a series of IJs near the bottom of a stack due to
inhomogeneous etching  around the boundary of a stack. The
irregularity in the values of $I_c$ and $I_r$ appearing in the
high-bias range in the initial etching stages is seen to
continuously reduce, and both of the quantities approach stable
values of their own in the low-bias range as $t_e$ increases [see
Figs. 4(b)-(f)]. This behavior implies that the branches with the
irregular $I_c$ and  $I_r$ correspond to  incompletely developed
junctions near the  bottom of the  stack. When the etching time is
increased beyond  10 min, at  least a few low-bias branches in the
$IV$ curves start showing almost the same critical currents ($I_c
\sim 2.3 \pm 0.1$ mA) and return currents ($I_r \sim 70 \pm 4
~\mu$A). Thus, a few small stacks are believed to develop fully
for $t_e \geq 10$ min. This conclusion is consistent with what we
discussed in relation with the $R_c(T)$ curves.

If we consider the stack as a $c$-axis  one-dimensional array of
the underdamped Josephson junctions,\cite{40,41,51} the McCumber
parameter\cite{52} $\beta_c$ deduced from the ratio\cite{53}
$I_r/I_c \sim 4/(\pi \sqrt{\beta_c}) \approx 0.031$ is about 1600,
which is about two  or three times larger than the results  of
others, $i.e.$, $\beta_c = 500$ (Ref. \onlinecite{11}) or
$300\sim700$ (Refs. \onlinecite{22a,24}). This difference may
arise from a high critical current density $j_c \simeq 1300$
A/cm$^2$ of our sample. One should note that the first
quasiparticle branch in the $IV$ curves in three-probe
measurements comes from the D$^{\prime}$ID surface WJ. The
magnified  view of the first branch in a very low-bias range in
the inset  of Fig. 4(b) shows a hysteretic behavior with a much
smaller critical current $I_c^{\prime}=46\pm 3 ~\mu$A and return
current $I_r^{\prime}=18\pm 1 ~\mu$A.

The negative dynamic resistance or the ``back-bending" for a set
of longer etching times in Figs. 4(d)-(f)   is a  generic  feature
in   the $IV$  characteristics of stacked   tunneling Josephson
junctions.\cite{54} Either a  nonequilibrium quasiparticle
distribution  in ultrathin conducting  layer\cite{3,6} or a Joule
heating effect\cite{55} has been proposed to explain the feature.
Fig. 4 clearly demonstrates that the back-bending develops
gradually as the number of IJs in the same stack increases,
attributing it to a high-bias  effect. In   Fig. 5 we   replotted
the high-bias   branches showing the back-bending behavior for the
different number of junctions in the stack. The back-bending
feature is visible, even from the early stages of etching
($t_e=15$ min corresponding to $n=6$), and becomes more pronounced
for longer etching.  Remarkable feature in this figure is that the
voltage-turning points  lie well  on the  3 mW power-dissipation
line.   Although no symptom  of temperature change was  detected
in  the thermometry,  this strongly  suggests that  Joule heating
could be a likely cause of the back-bending effect.

Careful examination of Fig. 5 reveals that as $t_e$ increases a
certain quasiparticle branch for a constant current bias shift  to
higher voltage range. This is opposite to  what one would expect
from the nonequilibrium quasiparticle distribution or  from the
Joule heating effect. We  infer that it arises from the increase
in the junction resistance $R_n^{\prime}$ and $R_n$ due to
shrinking of the junction area with increasing $t_e$.

In Figure 6, we replotted the first low-bias branch of $IV$ curve
for $t_e =6.5$ min which was already shown in Fig. 4(a) and the
first three low-bias branches of 6 sets of $IV$ curves
corresponding to   the etching times   from 10 min   to 30 min. We
also plotted  the anomalous branches of $IV$ curves for $t_e =3$
and 5 min. To eliminate the variation of the junction resistance
with $t_e$, the current axis in each $IV$ curve was rescaled by
multiplying $R_c(T_c)$ obtained from Fig. 2(a). One  first notices
that in this rescaled plot, with increasing $t_e$ up to 6.5 min,
the  over-all distribution of branches gradually  shifts to higher
voltages, and  the first branch of $t_e=6.5$ min almost coincides
with the first stable branch for $t_e=10$ min or longer. The
sequence of the branch  development indicates that the first
stable branch arises  from the surface WJ in D$^{\prime}$ID
configuration, and  the second  and third branches  from the inner
IJs,  both in  DID configuration.

If we assume that the D$^{\prime}$ID (DID) junction has the same
junction resistance $R_n^{\prime}$ ($R_n$) as the N$^{\prime}$ID
(NIN) junction, we can rewrite the $IV$ relation for the WJ and
IJs as follows.
\begin{eqnarray}\label{eq4}
  I_{D^{\prime}ID}\times R_n^{\prime}=\frac{1}{e} \int_{-\infty}^{\infty}
  N^{\prime}(E;\Delta^{\prime})N(E+eV;\Delta)[f(E)-f(E+eV)] dE
\end{eqnarray}
\begin{eqnarray}\label{eq5}
  I_{DID}\times R_n=\frac{1}{e} \int_{-\infty}^{\infty}
  N(E;\Delta)N(E+eV;\Delta)[f(E)-f(E+eV)] dE
\end{eqnarray}
in which $\Delta^{\prime}$ is the temperature-dependent suppressed
energy gap of the surface layer. Both $R_n^{\prime}$ and $R_n$
contain the possible variation of the junction area with etching
times. Eqs. (\ref{eq4}) and (\ref{eq5}) predict that each $IV$
curve of  the WJ or the  IJ will show  its own scaling behavior
with respect to  different junction resistances $R_n^{\prime}$ and
$R_n$, depending only on the energy gap $\Delta^{\prime}$ and
$\Delta$. The remarkable feature in Fig. 6 is that, when scaled by
$R_c(T_c)$ all together, all the  three branches, for
significantly different  etching times longer than 10  min, fall
into three merging curves. This confirms a definite
proportionality existing between $R_n^{\prime}$ and $R_n$,
regardless of the  etching time $t_e$. The point will  become
clearer in  the discussion in relation with Fig. 7(a) below.

We now determine the value of the zero-temperature  energy gap in
the surface layer, $\Delta_0^{\prime}$, by fitting the first
stable branch to Eq. (\ref{eq4}), while adopting $\Delta_0=32.6$
meV which was already obtained from the fit of $R _c(T)$ in Fig.
2(b) and setting $R_n^{\prime}$ to be $R_c(T_c)$ obtained from
Fig. 2(a). The best fit value\cite{56} is $\Delta_0^{\prime} =
15.3\pm 1.1$ meV assuming the BCS temperature dependence of gap
energy with $\alpha=1.45$. This value of the surface gap is  about
one-half of the intrinsic gap value  for any number of fully
developed junctions.  From the best fit of the second  or the
third  branch to Eq.  (\ref{eq5}), we obtain $R_n=1.54~\Omega$ for
$t_e =30$ min, which is about one sixth of $R_n^{\prime}$(30
min)$=R_c$($T_c$; 30 min). It yields the characteristic voltage
$I_c R_n \approx3.5$ mV, which is much smaller than $\pi \Delta_0
/2e =51.2$ mV predicted by the  Ambegaokar-Baratoff
relation\cite{57} in conventional superconductors  or the
theoretical prediction $\Delta_0/e$ for $d_{x^2 - y^2}$
symmetry.\cite{58} The value $R_n =1.54 ~\Omega$ corresponds to
$\rho_n \approx 23 ~\Omega$cm, which is close to the value of
$\rho_{c,single} \approx 20-22 ~\Omega$cm of IJs in the
high-temperature region (see Fig. 3). Thus the enhanced $R_c$
above $R_n$ around $T_{c,onset}$ is believed to be the
quasiparticle contribution discussed in Ref. \onlinecite{6}.

Since junctions in a stack switch to the higher resistive state
before reaching the gap edge of a junction, there are
uncertainties in determining both the normal-state resistance
$R_n$ and the critical current $I_c$ directly from the data. As
pointed out in Ref. \onlinecite{9}, in a fit to Eq. (\ref{eq5})
the values of $R_n$ and $\Delta_0$ are interrelated. For instance,
one gets smaller fit value of $R_n$ for the choice of larger
$\Delta_0$. We eliminated this uncertainty by adopting the
measured values of $R_c(T_c)$ for $R_n^{\prime}$, and
alternatively determined $\Delta_0$, $\Delta_0^{\prime}$, and
$R_n$, respectively, from the normalized $R_c(T)$ curves in
Region II  and $IV$  curves in Region I.  This process  is
possible  only for  repeated progressive measurements  in a single
piece  of crystal  with the same material  parameters
$\Delta_0^{\prime}$ and $\Delta_0$ through the entire
measurements.

The consistency of  the above fits  are further  ascertained by
cross-checking the etching time dependence of  the junction
parameters obtained by  different methods. In Fig. 7(a)  we plot
$R_n$ determined from a fit of the second branch to Eq.(\ref{eq5})
for different etching times of $t_e =10-30$ min. We also plotted
$R_c(T_c)$ obtained directly from Fig. 2(a) for different $t_e$.
Although showing a large difference in the absolute values for a
given $t_e$, the two data sets obtained by different ways, when
normalized by the values at $t_e =30$ min, $i.e.$, $R_c(T_c$; 30
min) $\equiv R_n^{\prime}$(30 min)$ = 9.52 ~\Omega$ and $R_n(30$
min$)=1.54 ~\Omega$, show an excellent agreement with one another
for all the ranges of $t_e$ values. This fact makes it clearer
that a definite proportionality exists between $R_n^{\prime}$ and
$R_n$ in any stack-forming etching stages. The value of $R_c(T_c)$
starts increasing for $t_e>5$ min and becomes fully  stabilized
for $t_e$ around  19 min.  The minimal  etching time required for
the full development of the stack can also be examined from the
$t_e$ dependence of the critical currents\cite{59} $I_c^{\prime}$
of the WJ and $I_c$ of IJs as shown in Fig. 7(b). The critical
currents were taken from the first ($I_c^{\prime}$) and the second
($I_c$) quasiparticle branches of the $IV$ curves in Fig. 4. Both
values of $I_c^{\prime}$ and  $I_c$ are stabilized for $t_e$
around $10-15$ min.

\section{Conclusions}
In this  study, we focused on the evolution of the tunneling
characteristics for stacks containing a few  intrinsic junctions,
for  temperatures above  and below the  bulk superconducting
transition temperature of  Bi-2212 single  crystals. Three-probe
measurements in comparison  with usual four-probe measurements
provided an  advantage of enabling us  to investigate the nature
of the surface junction with suppressed coupling  strength as well
as the  conducting properties of the inner junctions.   Compared
with   previous attempt  by others,\cite{8}  we were   capable of
lowering temperatures even below  the Josephson coupling
temperature  $T_c^{\prime}$ of  the surface  junction using
liquid-helium cooling. Our study reveals that $R_c(T)$ curves are
scaled by the junction resistance of the surface WJ below $T_c$.
On the other hand, the $IV$ charcteristics for both IJs and the WJ
in stable etching stage for $t_e \geq 6.5$ min are scaled by the
junction resistance of WJ, which indicates that a definite
proportionality exists between the values of junction resistances
of the WJ and IJs.

One of the main findings of this study is that the gap in the
surface layer $\Delta_0^{\prime}$ is much reduced from the bulk
value of IJs,  while the junction resistance of the  surface WJ
($R_n^{\prime}$) is much larger than that of the IJs ($R_n$). We
obtained $R_n$ from a fit to the second $IV$ branch as shown  in
Fig. 6. Mentioned earlier, in a fit to Eq. (\ref{eq5}) an almost
similar fit quality can be achieved by  choosing different values
for  interrelated parameters $R_n$ and $\Delta_0$. Instead of
obtaining $R_n^{\prime}$ from $R_c(T_c)$, just setting
$R_n^{\prime}$ to be $R_n$, for example, may be a reasonable
choice. With this choice, however, the fit gives
$\Delta_0^{\prime} \geq 3\Delta_0$, $i.e.$, the surface layer has
a  gap about three times larger than the  inner layers. Recently
it has been claimed that, in the underdoped regime, the gap
$\Delta_0^{\prime}$ has a larger value\cite{31,32,33} even for
lower transition temperature $T_c^{\prime}$. Thus it may seem
plausible to expect widening of the gap at the surface layer while
$T_c^{\prime}$ is suppressed. However, the universal relation
between the superconducting gap and the transition temperature in
Ref. \onlinecite{33} indicates that the magnitude of the gap for
$T_c^{\prime} \approx 30$ K is at best $\Delta_0^{\prime} \sim
1.3\Delta_0$, which is far smaller than $3\Delta_0$ obtained from
the fit above.  In addition, it  has also been  known that  the
suppression of  the transition temperature of Bi-2212 material
itself  down to $17-36$ K cannot  be achieved with a high degree
of oxygen deficiency, because a band with Bi-O  antibonding near
the Fermi level acts as a source of holes.\cite{60} The picture
that the  surface layer is more underdoped than the  inner layers
is again contradictory to the behavior revealed in Fig. 3,  where
the opposite was concluded earlier. In this sense, we discard the
picture of an increased gap with suppressed transition temperature
at the surface layer, and cling to the original results of fit
with a suppressed surface gap. In this case, however, we must
provide an explanation for the concurrent reduction of
$T_c^{\prime}$ and $\Delta_0^{\prime}$ in the surface layer, which
is apparently  in contradiction to the  known relation\cite{33}
between  the two variables.  The only  way to eliminate the
inconsistency is to  assume that  the concurrent suppression of
$\Delta_0^{\prime}$ and $T_c^{\prime}$ of the surface layer is not
caused  by the change in the doping level  in it, but by its
proximity contact to the Au normal electrode as supposed by us in
Ref. \onlinecite{11}. Recently, Manabe $et$ $al.$ observed that,
in scanning tunneling spectroscopy measurements on Bi-2212, the
gap becomes smaller as the tip approaches closer to the  crystal
surface.\cite{30} We believe this  is consistent with our claim of
the proximity-induced suppression of the gap on the surface layer
in our specimen. Although currently no theory is available to
allow  any quantitative confirmation of our  claim, this study
apparently strengthens our previous arguments.

Recently there have been tunneling  spectroscopic studies on the
nature of order-parameter symmetry of YBCO-123  and Bi-2212
superconducting  materials by  employing high-$T_c$
superconductor/normal metal/conventional $s$-wave superconductor
hybrid junctions.\cite{43,42} Studies have reported the  existence
of significant and  finite $s$-wave  components for  YBCO-123 and
Bi-2212 superconducting materials, respectively. Our study
indicates that in tunneling spectroscopy using the above-mentioned
hybrid junctions one should be concerned about the change in the
superconducting state at the crystal surface, not by  the oxygen
deficiency as one might  easily assume, but by the contact of
$d_{x^2 - y^2}$-wave  surface layers with a  normal layer in  a
junction. On the  other hand, STM tunneling spectroscopy may be
more affected by the oxygen deficiency at the surface.

The variation  of the   junction area along  with continuous
etching,  in  fact, acts  as  a disadvantage in studying the
inherent tunneling properties in  the system. Ironically, however,
this can be advantageous as in this study, because as illustrated
both  in Fig. 2 and in Fig. 6, we were able to be convinced that,
once the data are  normalized by proper area-dependent  parameters
like junction resistance,  the junctions  in any  etching stage
exhibit  no discernible variation in  their properties, regardless
of the number of junctions contained. This leads us to the
conclusion that not much correlation effect is present among the
IJs. Material parameters such as $\Delta_0^{\prime}$ and
$\Delta_0$ do not change in the whole in-situ etching/measurements
process, which  thus gives stronger conviction of the scaling
behavior in our $R_c(T)$ and $IV$ data as discussed above.

The in-situ technique is also extremely useful to study the
so-called back-bending effect. It convinces one that the effect
develops by increasing the number of junctions contained  in a
single piece of crystal, which in turn allows us to draw a
power-dissipation line. Comparing the back-bending feature in the
$IV$ branches from different crystals in terms of power
dissipation would be less meaningful. This technique  is expected
to  provide powerful  means to examine  both the interlayer
coupling properties in the superconducting state and the in-plane
conducting nature in the normal state.

\acknowledgments
We wish to acknowledge useful discussion with Prof. M. Tachiki.
We are grateful to Prof. M. Oda for providing valuable information
on his recent work on the doping-level dependence of the pseudogap
behavior.
This work was supported by the  Korea Science and  Engineering
Foundation under Contract No. K97005, Basic Science Research
Institute funded by the
Ministry of  Education and POSTECH under Contract No. 1NH9825602, and  the
Ministry of Defense through MARC.

\begin{figure}
\caption{ (a) An  optical micrograph of a sample prior to in-situ
ion-beam etching.  Four small stacks were fabricated on the top
surface of the base mesa.  A-F denote the Au electrodes. Data are
reported for the configuration where A and D (B and F) were used
for biasing currents and E and F (A and C) for voltage
measurements. (b) Enlarged schematic view of the dashed-circle
region in (a).}
\end{figure}

\begin{figure}
\caption{ (a) Progressive evolution of the $c$-axis resistance
$R_c(T)$ curves with increasing etching times; $t_e=0$, 3, 5, 6.5,
10, 12, 15, 19, 22, 26,  and 30 min from bottom to top. The
contact resistance was not subtracted. The characteristic
temperatures $T_c^{\prime}$, $T_{c,onset}^{\prime}$, $T_c$, and
$T_{c,onset}$ and  the temperature  regions I, II,  III are
defined  in the  text. The dotted lines show the $t_e$-dependence
of $T_c$ and $T_{c,onset}^{\prime}$. (b) A replot of the $R_c(T)$
curves, where $R_c(T)$ was rescaled by $R_c(T_c=86.5$ K$)$. The
dotted line is a fit to Eq. (\ref{eq2}) for a single
normal-metal/insulator/$d$-wave-superconductor (N$^{\prime}$ID)
surface junction with $\Delta_0 = 32.6$ meV, in comparison with a
fit (dashed line) with $s$-wave gap $\Delta_0 = 17$ meV. Inset: a
schematic view of the configuration for the in-situ three-terminal
measurements.}
\end{figure}

\begin{figure}
  \caption{
The single junction contribution to the tunneling resistivity
$\rho_{c,single}(T)$ as converted from the relation
$R_{c,single}(T)=[R_c(T,i+j)- R_c(T,i)]/j$ and the geometry
parameters for $i+j=8$, 9, 12, 12 and $i=6$, 6, 8, 6,
respectively. Inset: the $R_c(T)$ curves above $T_c$ for $t_e=15$,
19, 22, 26, and 30 min, corresponding to $n=6$, 8, 9, 10, and 12,
the total number of quasiparticle branches below $T_c^{\prime}$. }
\end{figure}

\begin{figure}
  \caption{
Current-voltage characteristics for the  etching time (a)
$t_e=6.5$ min ($n=4$), (b) 10 min ($n=5$),(c) 15  min ($n=6$),
(d) 19  min ($n=8$),  (e) 22  min ($n=9$), and (f)  30 min
($n=12$). The  substrate temperatures were in the  range
$13.9-16.8$ K. Inset  of (a): anomalous quasiparticle  branch
splitting for the etching of $t_e=3$ min and 5 min. Inset of (b):
the  enlarged view near zero bias which shows a hysteretic
behavior of  the surface WJ. Arrows indicate voltage changes
while varying the bias current. }
\end{figure}

\begin{figure}
  \caption{
A replot of the high-bias branches showing the back-bending
behavior in Fig. 4 for different numbers of junctions in the
stack. The dotted line represents the 3 mW power line, which
passes through the voltage-turning points of the  curves, pointing
toward the Joule-heating mechanism  for the back-bending behavior.
}
\end{figure}

\begin{figure}
  \caption{
Replot of the anomalous branches of $IV$ curves for $t_e =3$ (dot)
and 5 (bullet) minutes and the first low-bias branch of $IV$ curve
for $t_e =6.5$ (open circle) minutes shown in Fig. 4(a),  together
with the first three low-bias branches  of 6 sets of $IV$ curves
corresponding to the etching times of 10 (solid square), 12 (open
square), 19 (solid diamond), 22 (open diamond), 26 (solid
triangle), and 30 (open triangle) minutes. The dotted line in the
first  branch (the solid line in the second branch) is the fit to
the behavior of D$^{\prime}$ID  surface WJ (a DID inner IJ). The
specimen temperatures were  in the range $13.9-16.8$ K. }
\end{figure}

\begin{figure}
  \caption{
(a) The  etching time dependence  of $R_c(T_c)$ obtained from Fig.
2(a) and the normal-state resistance $R_n$ obtained from the
numerical fit  to Eq. (\ref{eq5}) of the $IV$  curves for the
second branches in  Fig. 6 below $T_c^{\prime}$.   For comparison
the  resistance values   were normalized by  those for   $t_e=30$
min; $R_c(T_c)=9.52 ~\Omega$ and $R_n =1.54 ~\Omega$. The lines
are guides to the  eyes. (b)  The etching time dependences of
$I_c$ and  $I_c^{\prime}$ at $T=13-16$ K. The critical currents
were taken from the first ($I_c^{\prime}$) and the second  ($I_c$)
quasiparticle branches of the $IV$ curves below $T_c^{\prime}$ as
shown in Fig. 4. The lines are guides to the eyes. }
\end{figure}


\begin{references}
\bibitem{46} The base mesa was consequently covered with $\sim$ 9000-\AA-thick Au film.
Deposition of a quite thick Au film was necessary to secure the electrical
connection across the crystal boundary.
\bibitem{47} This second superconducting transition is not shown in Refs.
\onlinecite{3} and \onlinecite{17}, where the specimens were annealed after gold
deposition.
\bibitem{56} In our previous work,\cite{11} we used the relation
$\Delta_0 ^{\prime}=\Delta{_0} T_c ^{\prime} / T_c$ to estimate $\Delta_0 ^{\prime}$.
Since the relation between the gap and the transition temperature due to proximity
contact of $d_{x^2 - y^2}$-wave superconductor to a normal metal is not known
we believe the direct extraction of the value from the fit as used in this study
has more relevance.
\bibitem{59} The critical current of the surface junction ${I_c}^{\prime}$,
due to its suppressed superconductivity, is always about an order of magnitude
smaller than those of inner junctions $I_c$.
\end{references}
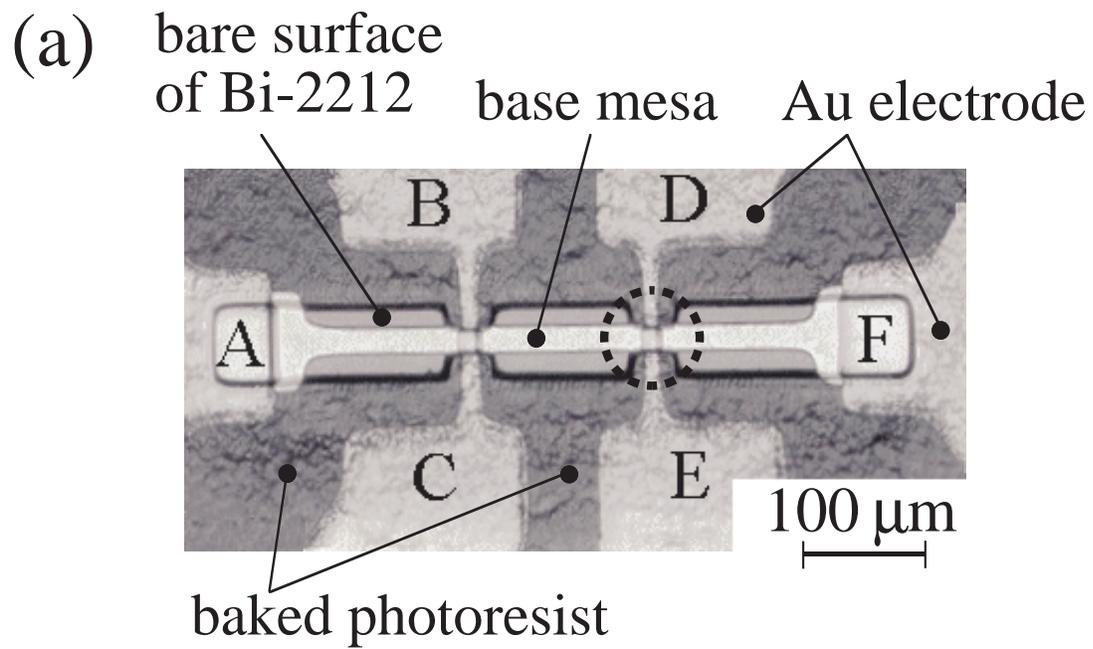
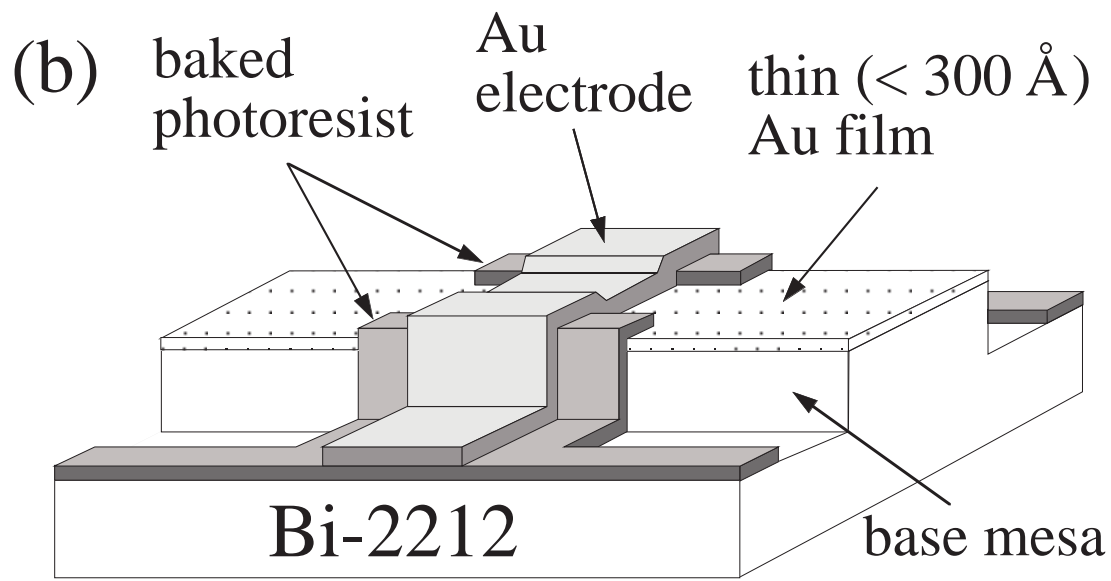
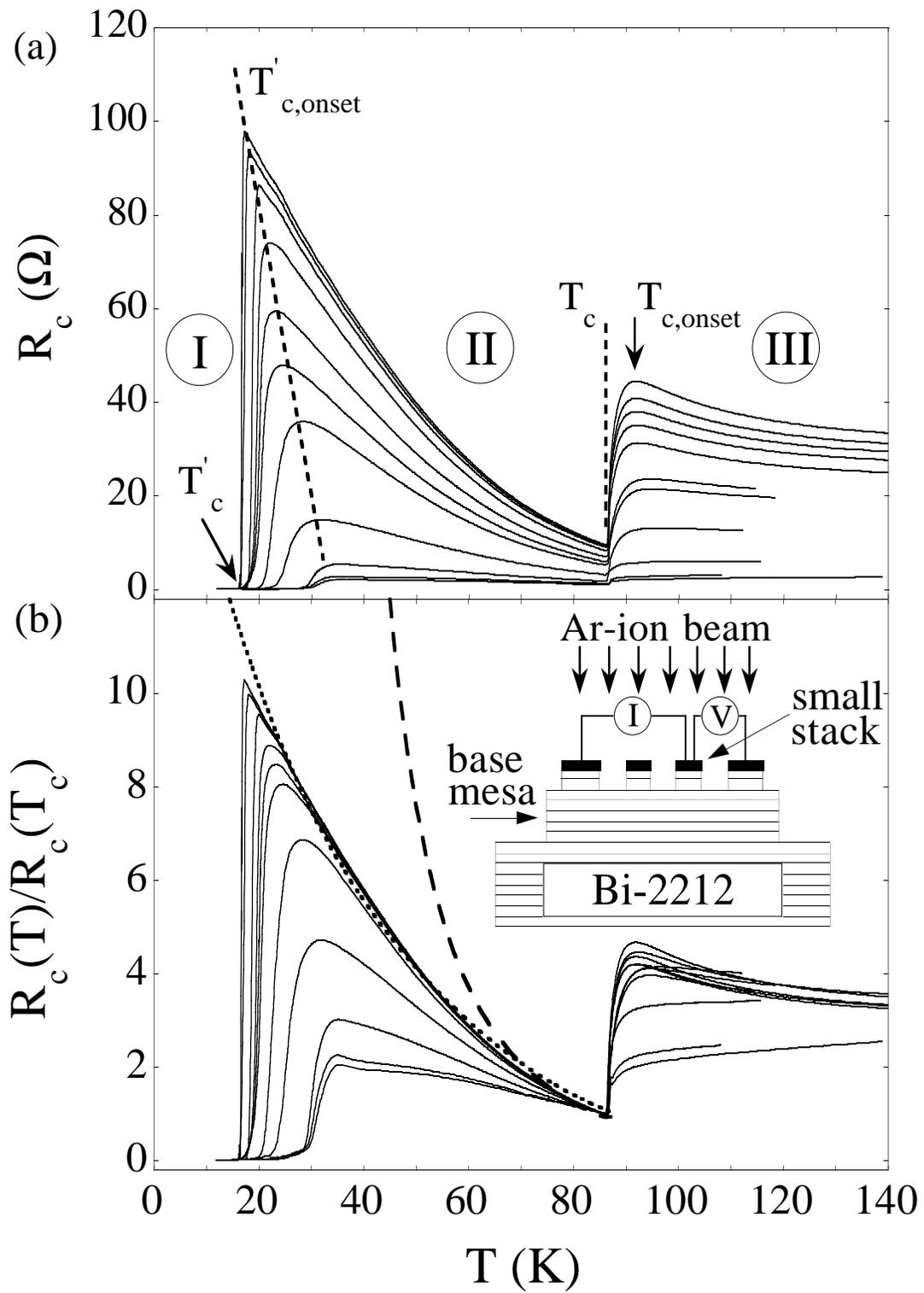
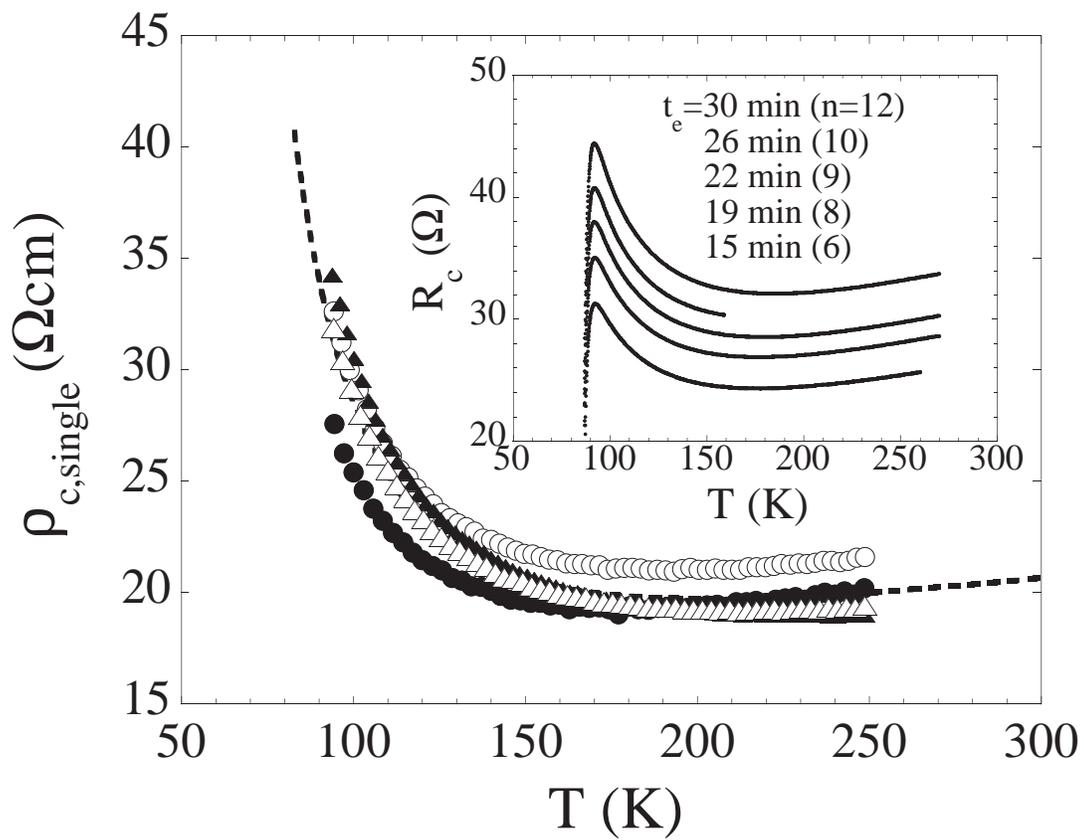
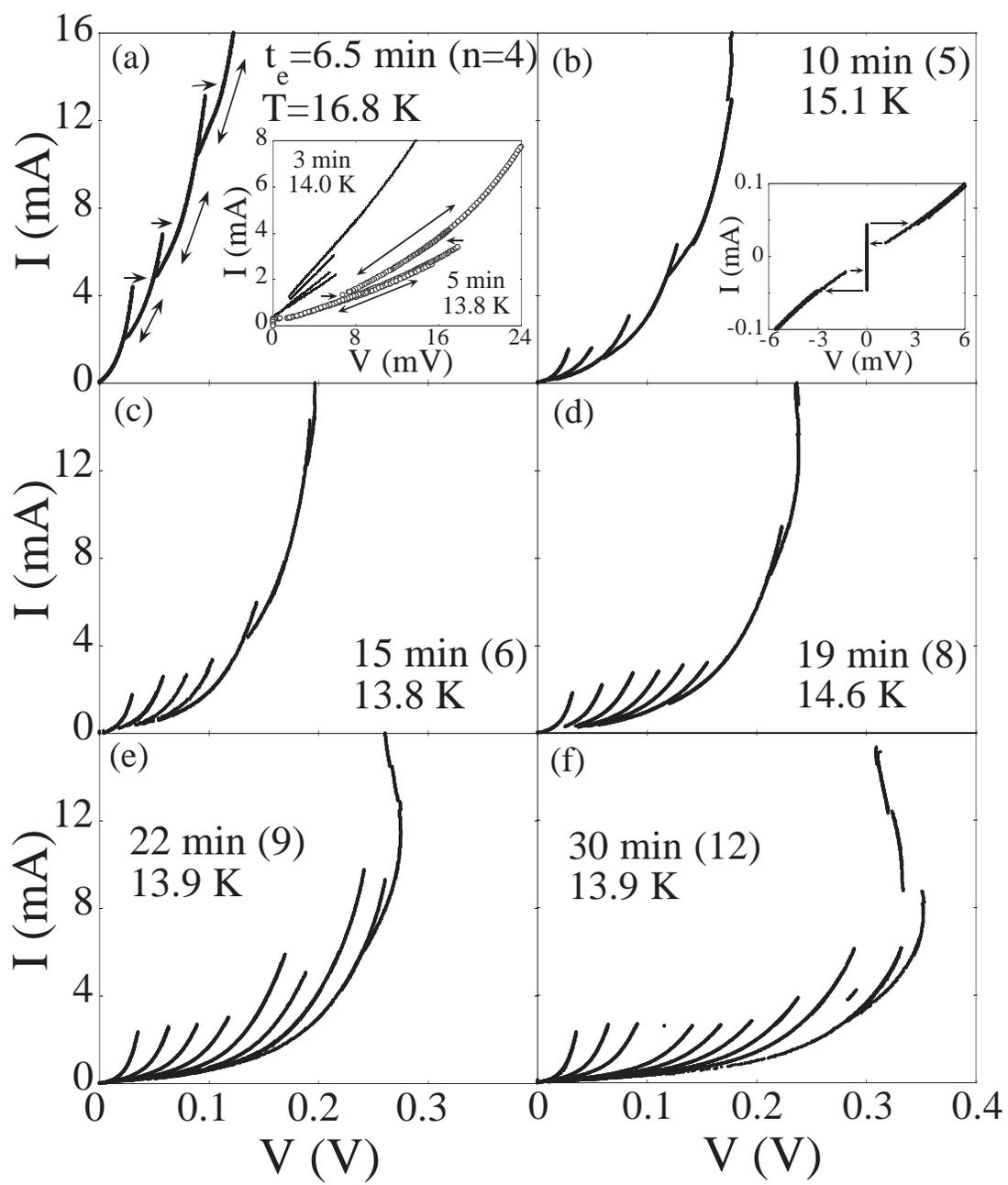
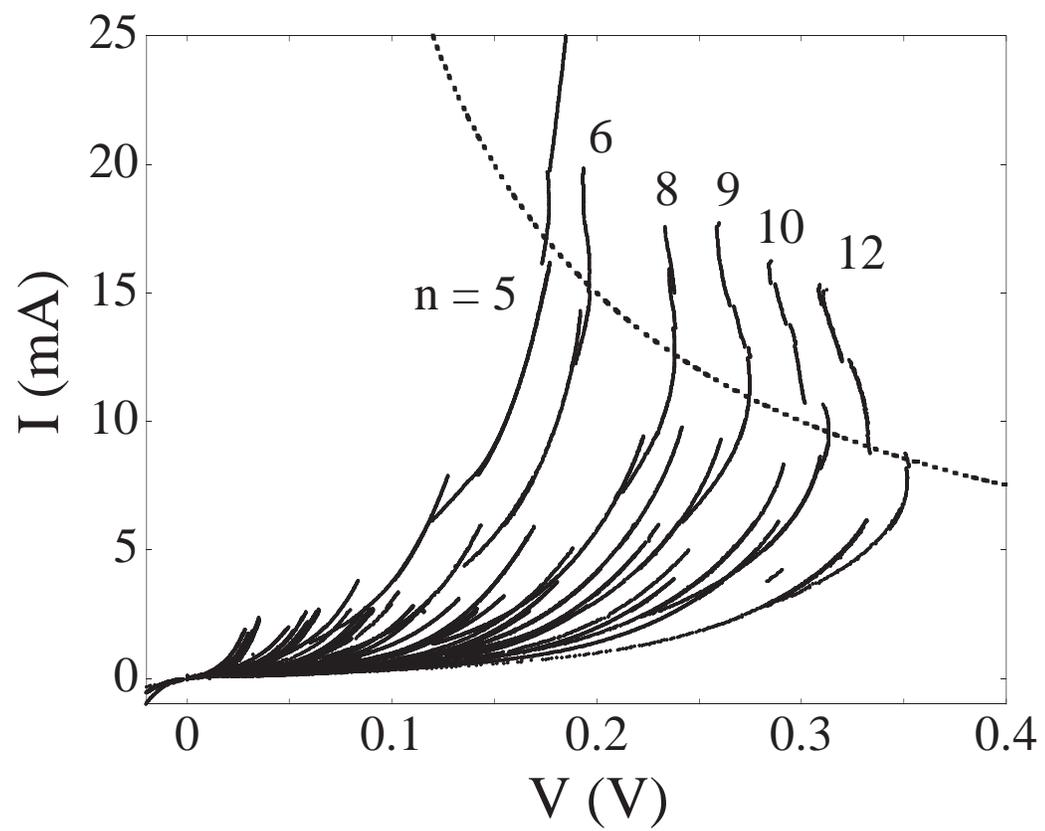
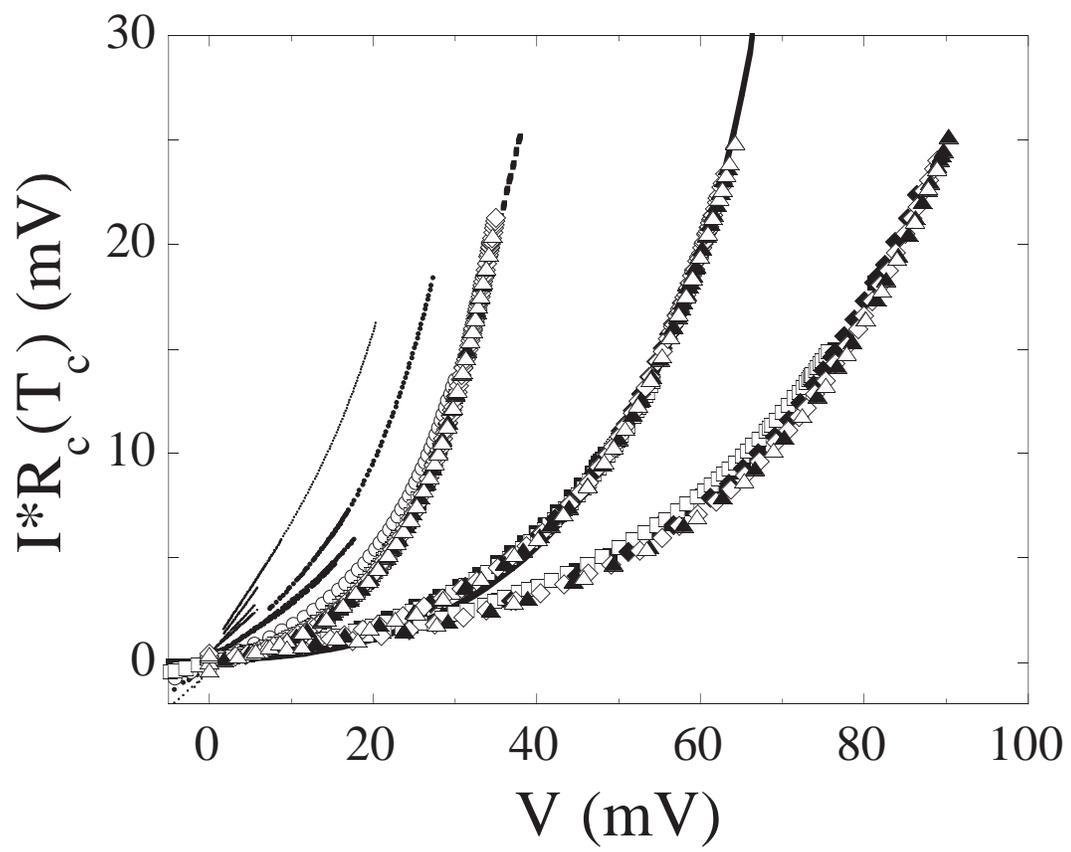
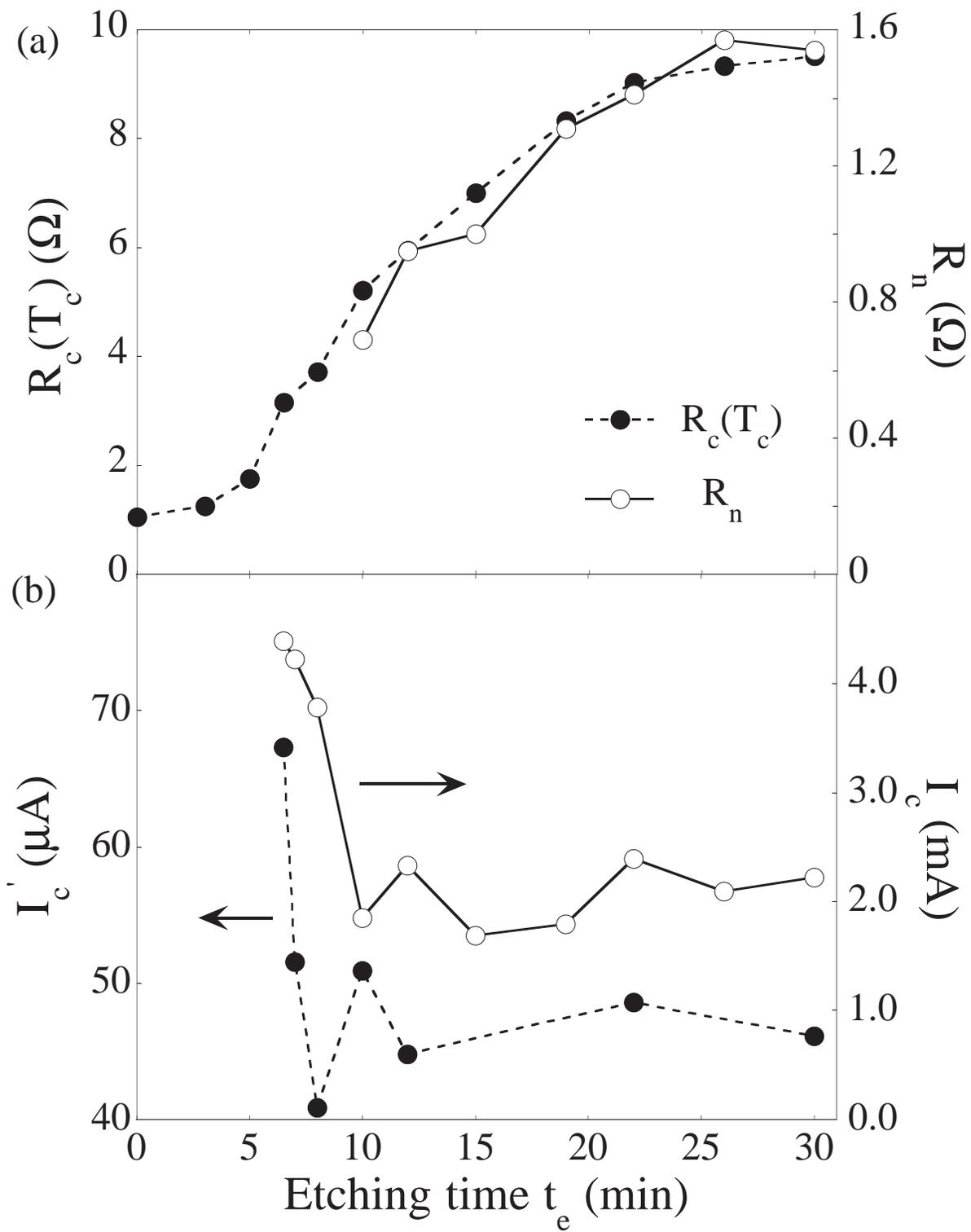
\end{document}